\documentclass[amsmath, amssymb, reprint, aip]{revtex4-1}
\usepackage{graphicx}
\usepackage{siunitx}

\begin{document}

\title{Limits on the Bolometric response of Graphene due to flicker noise}

\author{Sameer Grover}
\author{Sudipta Dubey}
\author{John P. Mathew}
\author{Mandar M. Deshmukh}
\email{deshmukh@tifr.res.in}
\affiliation{Department of Condensed Matter Physics and Materials
  Science, Tata Institute of Fundamental Research, Mumbai 400005, India}

\date{\today}

\begin{abstract}
We study the photoresponse of graphene field effect transistors using
scanning photocurrent microscopy in near and far field configurations,
and we find that the response of graphene under a source--drain bias
voltage away from the contacts is dominated by the bolometric effect
caused by laser induced heating. We find no significant change in the
photocurrent with the optical modulation frequency upto 100
kHz. Although the magnitude of the bolometric current scales with bias
voltage, it also results in noise. The frequency dependence of this
noise indicates that it has a 1/f character, scales with the bias
voltage and limits the detectable bolometric photoresponse at low
optical powers.
\end{abstract}

%\pacs{}

\maketitle

Graphene~\cite{Graphene_Science2004} --- an isolated single layer of
graphite --- promises to be useful for photodetection and other light
harvesting devices because of its gapless band structure and large
broadband absorption. Photodetectors made using
graphene~\cite{Avouris_Photodetector} involving global illumination
have been reported and the quantities of interest are the responsivity
and response time. In these measurements, the role of different
mechanisms involved in photogeneration is complex. In a recent
study,~\cite{CVDIR} the infrared photoresponse has been observed to
change with the graphene channel length suggesting different physical
mechanisms playing a role at the electrodes and far away from them.

Scanning photocurrent microscopy (SPCM) measurements are useful in
elucidating the mechanism of photocurrent generation in graphene since
the spatially resolved mapping of current can yield information about
the photoresponse at different regions on the graphene device such as
contacts and p-n junctions. Several
mechanisms~\cite{Graphene_SPCM_Review} have been identified such as
the photovoltaic,~\cite{Lee2008,JiwoongPark2009, Avouris_2009,
  Avouris_nearfield} photothermoelectric~\cite{MonoBiThermo,
  APL_thermoelectric} and bolometric effects.~\cite{Freitag_2012}

\begin{figure}[ht!]
\includegraphics[width=3.0in]{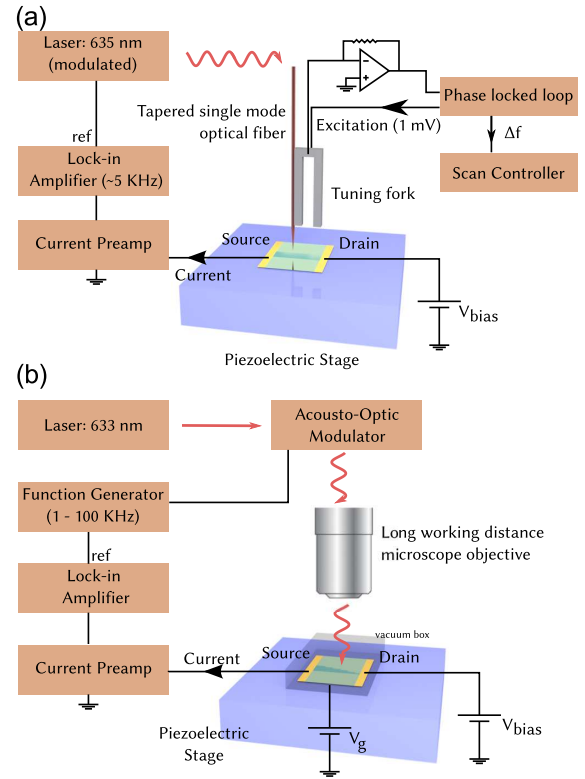}
\caption{\label{fig:schematic} Schematic of the (a) near--field and
  (b) far--field measurement setup. In (a), a tapered fiber is used
  for illumination (635 nm laser diode) as part of a home-built NSOM,
  and in (b) a commercial confocal Raman microscopy system is used
  with a 633 nm He-Ne laser. The graphene FET is mounted on a
  piezoelectric stage in ambient conditions in (a), and inside a
  vacuum chamber in (b). The AC current across the device is measured
  at the laser modulation frequency under the application of a DC gate
  and bias voltage. The AC current flowing in the direction of the DC
  current and in phase with the light modulation is considered
  positive.}
\end{figure}

\begin{figure*}
\includegraphics[width=5.7in]{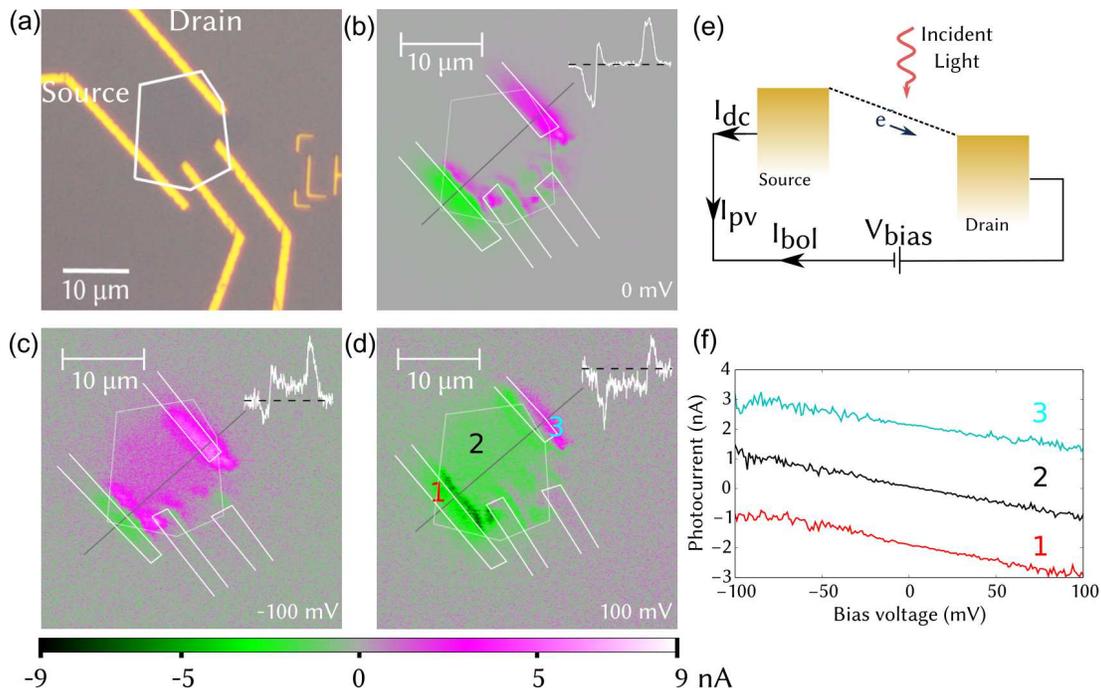}
\caption{\label{fig:nsom} Near-field photocurrent studies:(a) Optical
  image of graphene device D1. The edges of the graphene flake are
  shown. Also indicated are the source and drain electrodes. The other
  two electrodes are floating. Al$_2$O$_3$ dielectric of thickness 23
  nm was deposited using atomic layer deposition. (b), (c), (d)
  Near-field photocurrent maps of the in-phase component of the
  photocurrent at source--drain biases of 0, -100 and +100 mV
  respectively. The approximate position of the electrodes extracted
  from the simultaneously measured topographic AFM images are
  marked. A line slice of the data is also shown in the top right
  along with the zero photocurrent level. (e) Direction of the DC
  current, photovoltaic and bolometric contribution to the total
  current under an applied bias. (f) Variation of photocurrent as a
  function of applied bias at selected points in the scan window,
  points indicated in (d). An average optical power of 0.72 mW was
  coupled to the fiber.}
\end{figure*}

\begin{figure}
\includegraphics[width=3.1in]{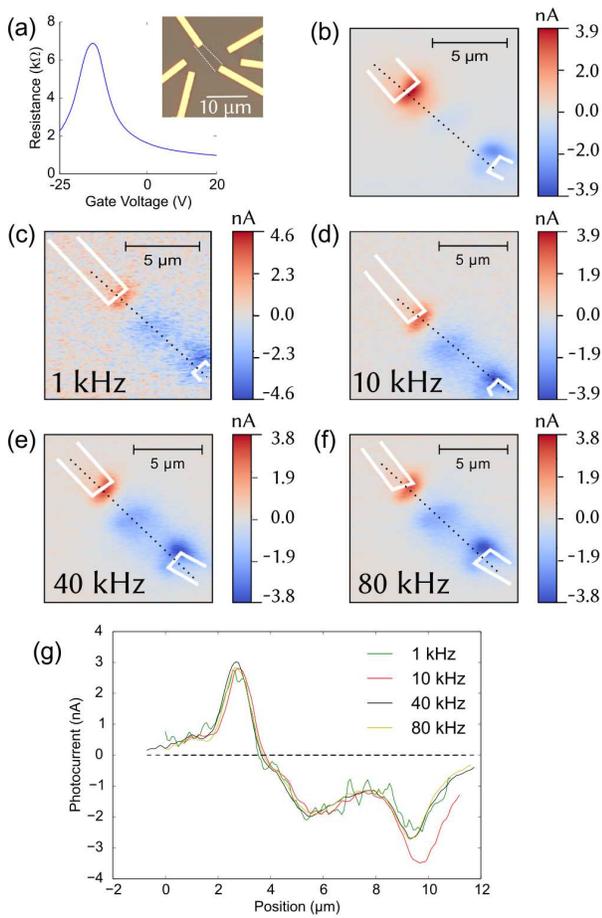}
\caption{\label{fig:freq}Photocurrent maps as a function of modulation
  frequency of light in far-field geometry (a) Optical microscope
  image of device D2 (boundary of graphene is outlined) and the gating
  curve. (b) Zero bias far field photocurrent map. (c), (d), (e) and
  (f) are photocurrent maps at a bias voltage of 80 mV and a gate
  voltage of 0 V with modulation frequencies of 1, 10, 40 and 80 kHz.
  The non-zero background has been subtracted. An average laser power
  of 40 $\mu$W is incident on the device. (g) Overlaid line
  slices along the indicated lines in (c)--(f) averaged over a width
  of 850 nm.}
\end{figure}

\begin{figure*}
\includegraphics[width=5.7in]{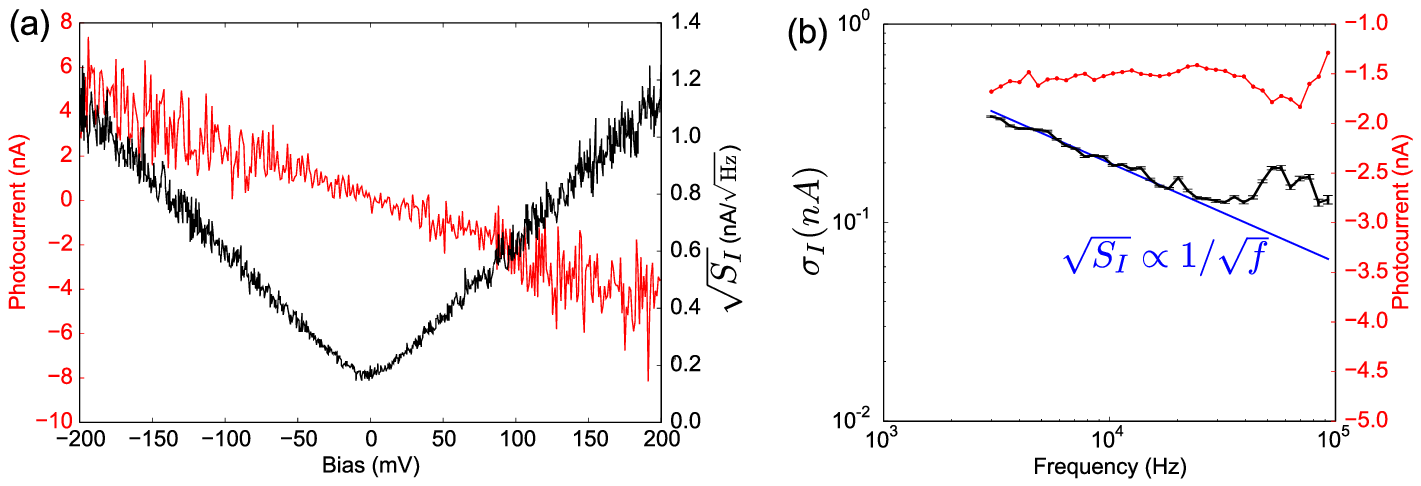}
\caption{\label{fig:freq2} Flicker noise in photocurrent measurements.
  (a) The photocurrent as a function of bias (red) at the center of
  the device at zero gate voltage. The square root of the current
  spectral density of the lock-in X-component (black) measured as a
  function of bias voltage at zero gate voltage. This measurement has
  been done at 1~kHz (b) The current fluctuation $\sigma_I$ as a
  function of frequency (black) calculated by sampling the lock-in
  R-component and fitting its distribution to a Gaussian. The width of
  this Gaussian has been plotted on the y-axis. This quantity is
  proportional to the square root of the current spectral density,
  i.e. $\sigma_I \propto \sqrt{S_I}$. The 1/f trendline is also
  shown. The change in the photocurrent as a function of frequency
  (red) at a bias voltage of 80 mV is calculated by subtracting the
  lock-in amplifier X-component measured in the center of the graphene
  sheet with the background value of the X-component.}
\end{figure*}

In this paper, we discuss SPCM measurements on graphene field effect
transistor (FET) devices using both near and far field configurations
and focus on homogeneous graphene away from the electrodes. The
photocurrent generation in graphene is dominated by the bolometric
effect far away from the electrodes and by the photovoltaic effect due
to the built-in electric field at the contacts. The bolometric effect
refers to the resistance change induced in the device by laser-induced
heating and it has previously been seen in carbon
nanotubes,~\cite{JiwoongCNT} graphene~\cite{Freitag_2012} and in 100
nm thick black phosphorus.~\cite{avouris_blackphosphorus} Our main
finding is that the bolometric effect, which is only visible at
non-zero bias voltages is accompanied by flicker noise which scales
with bias and limits the detectable bolometric signal.

Graphene devices were fabricated by exfoliation of graphite on
degenerately doped silicon substrates with 300 nm of silicon
dioxide. Monolayer graphene flakes were identified by visual contrast
in optical microscope and with Raman spectroscopy. Electron beam
lithography and thermal evaporation of Cr (10 nm)/Au (50 nm) were used
to define source and drain electrodes. We show measurements on two
devices in this paper: the first device (D1) has been used in
near-field measurements and the second device (D2) in a far-field
measurement setup. Measurements carried out on other devices are given
in the supplementary information.~\cite{supporting_info}

A schematic of the setup used is shown in Figure~\ref{fig:schematic}.
Light is modulated using an acousto-optic modulator (AOM) at
frequencies from 1 kHz up to 100 kHz. As a function of the applied DC
gate and bias (source--drain) voltage, the current induced across the
source is detected at the light modulation frequency using a lock-in
amplifier. The in-phase component of the AC current flowing in the
direction of the DC current has been taken to be positive
(Figure~\ref{fig:schematic}). The far field regime results in a
diffraction limited spot size of $\sim$ 1 $\mu$m in contrast with a
higher resolution of $\sim$ 250 nm using a near-field scanning optical
microscope (NSOM).

Figure~\ref{fig:nsom} shows the near field SPCM measurements performed
on D1. In these measurements, we observe a non-zero background current
that is measured when the light is incident on the oxide away from the
graphene flake. This has previously been attributed to the light
induced photovoltage~\cite{Freitag_2012} produced at the
silicon-silicon dioxide interface which gates the graphene at the
modulation frequency and results in a finite measured photocurrent
even when the laser is not falling directly on the graphene flake. We
compensate for this by subtracting out a constant value from the
photocurrent maps so as to make the average background value
zero. Strong localized photocurrent is produced near all the
electrodes with a large magnitude near the source and drain and a much
smaller magnitude near the floating electrodes. We attribute this to
the drift currents produced by built--in electric field at the
metal--graphene interface because of the difference in work
functions. These features have been analyzed in detail in previous
work.~\cite{JiwoongPark2009} Thermoelectric contributions, which may
also play a role, can be distinguished from photovoltaic contributions
by changing the polarization of light.~\cite{NovoselovPolarization}

Furthermore, after correcting for the background, at zero bias, there
is no current being produced in the graphene region far from the
electrodes, but at high biases, there is uniform photocurrent
generation throughout graphene. Photocurrent due to the photovoltaic
effect and the tilting of the bands is expected to be in the direction
of DC current (Figure~\ref{fig:nsom}(e)). The observed direction of
current is opposite this and therefore it cannot be attributed to the
photovoltaic effect. This is instead due to the bolometric effect in
which the laser induced heating of graphene causes the resistance to
increase, which, in turn causes the photocurrent to flow in a
direction opposite to the DC current flowing through the device.

We observe that the data gets noisier at large bias voltages. This is
evident in Figure~\ref{fig:nsom} where it can be seen that the
background in the photocurrent maps at -100 mV and +100 mV is
noisier. Similar results have also been reported in other scanning
photocurrent studies but this has not been examined in detail; for
example, with MoS$_2$.~\cite{NoisyMoS2} This noise is not related to
the measurement circuit but is intrinsic to the system. It is absent
at zero bias and is seen to scale with the bias voltage. Most previous
SPCM measurements on graphene have been done at zero bias where this
noise is not visible.

In order to understand the origin and frequency dependence of the
noise, we have measured photocurrent maps at different light
modulation frequencies. These measurements were done on D2 in a far
field system. Our results are summarized in Figure~\ref{fig:freq}
which shows SPCM measurements at a bias of 80 mV and gate voltage of
0~V at modulation frequencies of 1, 10, 40 and 80 kHz. We can observe
a strong dependence of the noise on the frequency, with the data
noisier at lower frequencies.

We have also varied the bias voltages and frequency while keeping the
laser at one point. The results for device D1 in which the bias
voltage is scanned from -100 mV to +100 mV are shown in
Figure~\ref{fig:nsom}(f) after compensating for the background. It can be
seen that in the center of the graphene flake, there is no
photocurrent but as the bias voltage is increased, the magnitude of
the photocurrent increases linearly. The negative slope is indicative
of the bolometric effect. In the bolometric effect, a change in
temperature $\Delta T$ causes a change in the conductance $\Delta G =
(dG/dT) \Delta T$ which in turn leads to an AC current $\Delta I =
V_{bias} \times \Delta G = V_{bias} (dG/dT) \Delta T$. For graphene
close to room temperature, $dG/dT < 0$,~\cite{Freitag_2012} we get a
negative current.

An average laser power of 40 $\mu$W has been used for device D2. Using
this value, we can estimate the change in temperature of graphene
using Fourier's law. Assuming the heat spreads radially around the
light spot, we can estimate the temperature
increase~\cite{Balandin_2008} as $\Delta T = \frac{\alpha P}{2\pi h k}
= 783$~$\textrm{mK}$ where $h$ is the thickness of monolayer graphene,
$P$ is the incident optical power, $\alpha$ is the fraction of power
absorbed by the graphene sheet and $k = 5.5\times10^2$ $\textrm{W
  m}^{-1}\textrm{K}^{-1}$ is the in-plane thermal conductivity for
supported graphene.~\cite{MRS}  We have also solved the classical heat
equation in two dimensions analytically and using the finite element
method (details in supplementary information~\cite{supporting_info}) and obtain a temperature
increase of 800 and 200 mK respectively. The change in conductance
$|\Delta G|$ at room temperature, as measured in \cite{Freitag_2012}
is approximately $50$ $\textrm{nS/K}$. Using a temperature increase of
500 mK, we get a conductance change of $|\Delta G|=25$ nS. The
corresponding experimental value is 10 nS for device D1 and 20 nS for
device D2.

The photocurrent as a function of bias voltage and frequency at the
center of the device D2 are shown in Figure~\ref{fig:freq2}. The
difference of the average value of the fluctuating in-phase component
in the center of the graphene flake and the background (Si/SiO$_2$)
gives us the change in photocurrent as a function of frequency. We
have observed changes in the photocurrent maps between scans with the
same parameters, such as forward and backward scans done with the near
field (see also Figure~\ref{fig:freq}(g)) and the changes shown in
Figure~\ref{fig:freq2}(b) of the order of 600 pA are not
significant. The bolometric response arises due to a local increase in
temperature. The net temperature rise is due to a balance between
escape of heat due to large thermal conduction, and absorption due to
finite specific heat. We expect a change~\cite{BolometerReview} in the
current at time scales of $\tau \sim A/\alpha_{T}$ where the the area
of graphene $A = 20$ $\mathrm{\mu m^2}$ and the thermal diffusivity
$\alpha_{T} = k/\rho C = 3.8 \times 10^{-4}$ m$^2$/s ($k$: thermal
conductivity, $\rho$: density, $C$: specific heat). This evaluates to
$1/\tau \sim 20$ MHz.

We have also studied the characteristics of the noise by measuring its
dependence on frequency and bias voltage. Photocurrent measurements at
different modulation frequencies in device D2
(Figure~\ref{fig:freq}(c)-(f)) indicate that the noise reduces as the
frequency is increased. To calculate the current spectral density
$S_I$, we sweep the light modulation frequency from 3~kHz to 100~kHz
and at each frequency, the R component of the lock-in amplifier is
sampled and recorded as a function of time. The histogram of the R
values at each frequency fits well to a Gaussian distribution and its
width is a measure of the temporal fluctuation in photocurrent. We
have called this quantity the current fluctuation $\sigma_I$ (shown in
Figure~\ref{fig:freq2}(b)). This is related to the root of the current
spectral density by a factor equal to the square root of the
equivalent noise bandwidth ($B$) as $\sqrt{S_I} = \sigma_I/\sqrt{B}$.
We carry out this measurement with the laser spot positioned in the
centre of graphene to quantify the noise in the photocurrent. This is
plotted in Figure~\ref{fig:freq2}(b) and is seen to be in good
agreement with a $\sqrt{S_I} \propto 1/\sqrt{f}$, or equivalently $S_I
\propto 1/f$ dependence as seen in other graphene
devices.~\cite{balandin_noise} We can further quantify the
noise~\cite{VandammePANI,VandammeCircular} using
Figure~\ref{fig:freq2} as $S_I/I^2 = C_{1/f}/f$ and obtain $C_{1/f} =
7.2 \times 10^{-8}$. The noise normalized to frequency and area,
$C_{us} = C_{1/f} \times \mathrm{Area}$ is $C_{us} = 1.2 \times
10^{-6} \mu m^2$. Hooge's parameter\cite{Hooge,Hooge2,VandammeHooge}
$\alpha_H = C_{1/f} \times N$ where $N$ is the total number of charge
carriers is $\alpha_H = 1.4\times 10^{-2}$.

We have also studied the dependence of the noise on the bias and
Figure~\ref{fig:freq2}(a) shows that the scaling of the noise on bias is
in accordance with Hooge's empirical law, with the spectral density
$\sqrt{S_I} \propto V_{bias}$. The dependence of the photocurrent on
the optical power (supplementary information~\cite{supporting_info}
section V) shows that the signal to noise ratio improves with
increasing laser power. The photocurrent scales with both the optical
power and bias, and the noise scales with the bias voltage, indicating
that at low optical powers, the bolometric signal will not be
visible. Furthermore, the flicker noise will also contribute to the
non-zero background as any frequency--dependent AC components of the
noise will be picked up by the lock-in amplifier and will contribute
to the detected photocurrent. The current spectral density of the
non-zero background current measured when the laser is positioned on
the silicon dioxide substrate shows a 1/f trend identical to the one
shown in Figure~\ref{fig:freq2}(b). Increasing the time constant of the
lock-in amplifier will reduce the noise bandwidth.

There have been proposals of using graphene as a photodetector using
the bolometric effect~\cite{Freitag_2012} which has the advantage of
broadband detection.~\cite{terahertz} We have shown that there are
noise fluctuations in the photocurrent signal that scale with
frequency as 1/f. The bolometric effect, therefore, is accompanied by
flicker noise which limits the minimum detectable optical power
incident on graphene. We have estimated the noise equivalent power
(NEP) of the graphene bolometric detector in the flicker noise regime
as $1.6\times10^{-4}/\sqrt{f}$ W Hz$^{-1/2}$ valid for the operating
frequency $f < 40$ $\mathrm{kHz}$. Typical values of NEP for
commercial detectors are in the range~\cite{FibreOpticsBook} of 1 --
10 pW Hz$^{-1/2}$. Flicker noise in graphene arises from the
fluctuation in the number of charge carriers and their
mobility. Techniques to reduce it such as electron
irradiation~\cite{noisereduction, noise_reduction2, noise_reduction3,
  noise_reduction4, noise_reduction5} can be explored in this
context. In addition, the high NEP in graphene is also due to the low
sensitivity arising in part because of the weak dependence of the
graphene resistance with temperature.~\cite{Koppens2014}

In conclusion, we have studied the photoresponse of biased graphene
and observed that the bolometric effect dominates the
photoresponse. The bolometric effect is evident only at non-zero bias
voltages and scales linearly with the bias voltage. It is accompanied
by flicker noise which increases quadratically with the bias voltage
and reduces the signal to noise ratio in graphene photodetectors. We
have shown that an optical power of 40 $\mu$W which corresponds to a
temperature increase of $\sim$ 500 mK is sufficient to resolve the
bolometric contribution at bias voltages of $\sim$ 100 mV.

%\begin{acknowledgments}
  The authors thank Ravitej Uppu and N Sreeman Kumar from the Nano
  Optics and Mesoscopic Optics Laboratory, DNAP, TIFR for their
  assistance in fabrication of NSOM tips. This work was supported by
  the Swarnajayanti Fellowship from the Department of Science and
  Technology and the Department of Atomic Energy of the Government of
  India.
%\end{acknowledgments}

%

\end{document}